\documentclass[aps,prb,showpacs,noshowkeys]{revtex4}
\usepackage{graphicx}
\usepackage{epstopdf}
\usepackage{latexsym}
\usepackage{amssymb}
\usepackage{amsmath}
\usepackage{amsfonts}
\usepackage{subfigure}
\usepackage{bm}
\usepackage{multirow}

\begin{document}

\title{Mott physics, sign structure, ground state wavefunction, and high-T$%
_{c}$ superconductivity}
\author{Zheng-Yu Weng}
\affiliation{\textit{Institute for Advanced Study, Tsinghua University, Beijing 100084,
China}}

\begin{abstract}
In this article I give a pedagogical illustration of why the essential
problem of high-T$_{c}$ superconductivity in the cuprates is about how an
antiferromagnetically ordered state can be turned into a short-range state
by doping. I will start with half-filling where the antiferromagnetic ground
state is accurately described by the Liang-Doucot-Anderson (LDA)
wavefunction. Here the effect of the Fermi statistics becomes completely
irrelevant due to the no double occupancy constraint. Upon doping, the
statistical signs reemerge, albeit much reduced as compared to the original
Fermi statistical signs. By precisely incorporating this altered statistical
sign structure at finite doping, the LDA ground state can be recast into a
short-range antiferromagnetic state. Superconducting phase coherence arises
after the spin correlations become short-ranged, and the superconducting
phase transition is controlled by spin excitations. I will stress that the
pseudogap phenomenon naturally emerges as a crossover between the
antiferromagnetic and superconducting phases. As a characteristic of non
Fermi liquid, the mutual statistical interaction between the spin and charge
degrees of freedom will reach a maximum in a high-temperature
\textquotedblleft strange metal phase\textquotedblright\ of the doped Mott
insulator.
\end{abstract}

\date{{\small \today}}
\maketitle

\section{Introduction}

Right after the 1986 discovery of high-$T_{c}$ superconductivity in the
cuprate materials, Anderson made a seminal proposal \cite{pwa_87} that the
on-site Coulomb repulsion plays a crucial role. At half-filling, with each
unit cell in the copper-oxide plane occupied by one electron from the
\textquotedblleft conduction band\textquotedblright , the cuprate is a Mott
insulator with a full gap opening up in the charge degree of freedom. Due to
the odd number of $S=1/2$ spin per unit cell, the spin degrees of freedom
remain unfrozen, in contrast to a band insulator, and the localized spins
antiferromagnetically interact with each other via a residual
nearest-neighbor Heisenberg superexchange coupling $J$, to the leading order
of approximation. It was then conjectured \cite{pwa_87} that the novel
superconductivity should arise by doping such a Mott insulator.

A prototype superconducting ground state incorporating the \textquotedblleft
Mottness\textquotedblright\ is known as the Gutzwiller-projected BCS state
\cite{pwa_87,pwa_03,gros_07}%
\begin{equation}
|\Psi _{\mathrm{RVB}}\rangle =\hat{P}_{\mathrm{G}}|\mathrm{BCS}\rangle
\label{BCS}
\end{equation}%
where $|\mathrm{BCS}\rangle $ denotes an ordinary BCS superconducting state
and $\hat{P}_{\mathrm{G}}$ is a Gutzwiller projection operator enforcing the
following no double occupancy constraint

\begin{equation}
\sum_{\sigma }c_{i\sigma }^{\dagger }c_{i\sigma }\leq 1.  \label{mottness}
\end{equation}%
There are two essential components in this ground state ansatz. First, the
Cooper pairing in $|\mathrm{BCS}\rangle $ is originated from the
superexchange coupling $J$. Second, the Gutzwiller projection $\hat{P}_{%
\mathrm{G}}$ removes those doubly occupied configurations which would cost
high energy due to the large on-site Coulomb repulsion. In particular,
because of $\hat{P}_{\mathrm{G}}$, the Cooper pairing in $|\mathrm{BCS}%
\rangle $ reduces to the neutralized singlet pairs of spins at half-filling,
known as the resonating valence bond (RVB) state \cite{pwa_87}, which is a
Mott insulator with the charge degree of freedom totally frozen out.

Hence an essential observation made in Ref. \cite{pwa_87} is that the
localized spins have a tendency to form real-space singlet pairing in order
to gain the superexchange energy. Hole doping (with removing the localized
spins) can then lead to high-$T_{c}$ superconductivity as the charge neutral
RVB pairs start to move to the empty sites and become partially charged
Cooper pairs. Today, strongly motivated by this idea, searching for new Mott
insulators with an RVB (spin liquid) ground state has become an important
goal in material syntheses, with the hope to find new exotic superconductors
upon doping.

However, the electron ground state in the half-filled cuprates has turned
out to be not simply a spin liquid with short-range RVB pairing. As a matter
of fact, it was quickly established experimentally \cite{exp1} that the
localized spins are long-range antiferromagnetically (AFM) ordered instead.
Theoretically the ground state of the AFM Heisenberg model in
two-dimensional (2D) square lattice is also an AFM ordered state \cite%
{theory} whose properties are well in accord with those found in the
cuprates at half-filling. Nevertheless the AFM long-range order disappears
quickly upon doping the holes into the cuprates. Consequently only
short-range AFM correlations remain beyond some finite doping concentration,
which may be still effectively described, as one would hope, by some kind of
RVB or spin liquid state \cite{LNW_06}. Since there is already a sufficient
amount of holes present, it would then become superconducting as envisaged
\cite{pwa_87} in the original RVB proposal.

In this context, a spin liquid state relevant to the superconductivity, if
present in the cuprates, should be by itself an \emph{emergent} state,
arising after the long-range AFM order of an antiferromagnet gets destroyed
by the motion of the doped holes. The real challenge to understanding the
superconducting mechanism in the cuprates is therefore to correctly
demonstrate how the antiferromagnet at half-filling can be doped into a
short-range AFM state, where the doping itself plays the central role of
\textquotedblleft dynamic frustration effect\textquotedblright . The
resulting state can be either simultaneously superconducting or, more
generally, a \textquotedblleft pseudogap phase\textquotedblright\ with the
superconducting phase naturally embedded inside.

This will be the central theme illustrated in this article. I will present
convincing theoretical rationale of why it is absolutely necessary to
carefully deal with the non-perturbative and singular nature of doping an
antiferromagnet in order to correctly understand the superconducting and
pseudogap physics in the cuprates. The resulting superconducting ground
state \cite{weng_11}, although bears a close resemblance to Eq. (\ref{BCS}),
is generally qualitatively different from the latter because, as it turns
out, the spin and charge degrees of freedom are intrinsically entangled
together by a hidden fundamental law underlying the Mottness (see Sec. III),
which is not manifested in Eq. (\ref{BCS}).

\section{Constructing a new superconducting ground state}

As emphasized in Introduction, the superconducting ground state in a doped
Mott insulator will reduce to a Mott insulator at half-filling. In
particular, the localized spin state should correctly recover the AFM
long-range ordered ground state of the Heisenberg model in this limit. Below
we illustrate how to realize such a construction in a straightforward way.
\begin{figure}[tbph]
\begin{center}
\includegraphics[width=130pt]{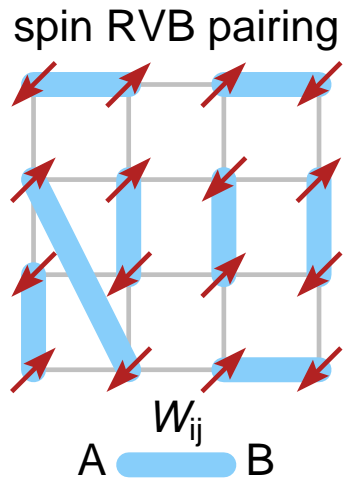}
\end{center}
\caption{In an LDA wavefunction (Eq. (\protect\ref{RVB})), bosonic spins
form singlet pairs with an amplitude $W_{ij}$, which only connects two spin
partners at opposite (A and B) sublattices as indicated by a blue-colored
bond. The Marshall sign rule \protect\cite{marshall_55} is precisely
satisfied at half-filling, where $W_{ij}\propto 1/\left\vert \mathbf{r}%
_{ij}\right\vert ^{3}$ is real and positive, and the wavefunction is very
accurate to describe the ground state of the Heisenberg model with a
long-range AFM order in the thermodynamic limit \protect\cite{lda_88}. In
the superconducting phase, such a state will reduce to a \textquotedblleft
ghost\textquotedblright\ spin liquid background with $W_{ij}$ becoming
short-ranged (cf. Fig. \protect\ref{wf}). }
\label{b-RVB}
\end{figure}

At half-filling, although there is no exact solution in 2D, the
Liang-Docout-Anderson (LDA) type bosonic RVB state \cite{lda_88} can give
rise to the most accurate variational energy for the Heisenberg model with
the correct long-range AFM order. It can be written in the form%
\begin{equation}
|\mathrm{RVB}\rangle =\sum_{\{\sigma _{s}\}}\Phi _{\mathrm{RVB}}\left(
\sigma _{1},\sigma _{1},\cdot \cdot \cdot ,\sigma _{N}\right) c_{1\sigma
_{1}}^{\dagger }c_{2\sigma _{2}}^{\dagger }\cdot \cdot \cdot c_{N\sigma
_{N}}^{\dagger }|0\rangle  \label{RVB}
\end{equation}%
in the electron $c$-operator representation. By nature this is a \emph{%
bosonic} state with a bosonic wavefunction $\Phi _{\mathrm{RVB}}\left(
\{\sigma _{s}\}\right) \equiv \sum_{\mathrm{partition}%
}\prod_{(ij)}(-1)^{i}W_{ij}$ for each given spin configuration $\{\sigma
_{s}\}=\sigma _{1},\sigma _{1},\cdot \cdot \cdot ,\sigma _{N}$ as shown in
Fig. \ref{b-RVB}$.$ Here the RVB pairing amplitude $(-1)^{i}W_{ij}$ connects
two \emph{antiparallel} spins denoted by $i$ (up spin) and $j$ (down spin)
at two opposite sublattices of the square lattice, with the summation
running over all possible pairing partitions for a given $\{\sigma _{s}\}$.
The staggered sign $(-1)^{i}$ (the Marshall sign) is explicitly separated
from $W_{ij}$ such that the latter remains a smooth function of the distance
between even and odd lattice sites. A real and positive $W_{ij}$ obeying the
power law at large spatial separation of $ij$ has been shown to optimize
\cite{lda_88} the ground state energy, with $\Phi _{\mathrm{RVB}}$
satisfying the Marshall sign rule \cite{marshall_55}.

\begin{figure}[bp]
\begin{center}
\subfigure[]{\includegraphics[width=100pt]{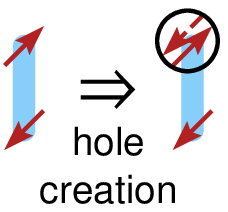}}\qquad\qquad\qquad %
\subfigure[]{\includegraphics[width=130pt]{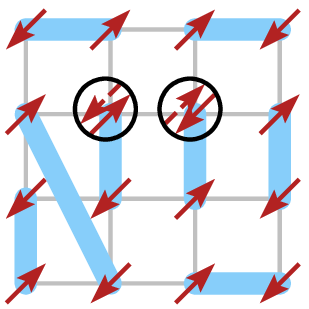}}\qquad\qquad\qquad %
\subfigure[]{\includegraphics[width=130pt]{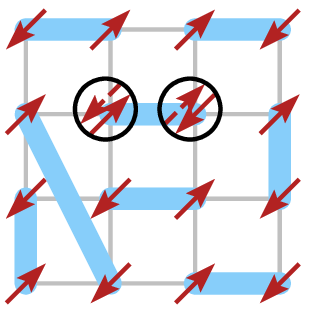}}
\end{center}
\caption{Schematic illustration of the new superconducting ground state (%
\protect\ref{gs}). (a) An RVB pair in Fig. \protect\ref{b-RVB} becomes a
hole-spin pair with the hole (the circle) created by annihilating a spin (as
indicated by the dashed arrow), with its RVB partner being automatically
associated with the doped hole; (b) and (c) Two doped holes moving on $|%
\mathrm{RVB}\rangle $ have the tendency to form a Cooper pair as driven by
the RVB pairing between their spin partners. Note that the important sign
structure of the wavefunction, i.e., $\Lambda _{h}$ in Eq. (\protect\ref{gs}%
), is not directly shown here.}
\label{wf}
\end{figure}

To start with $|\mathrm{RVB}\rangle $, which is much more precise than Eq. (%
\ref{BCS}) at half-filling, a ground state ansatz for the doped case may be
constructed as follows \cite{weng_11}%
\begin{equation}
|\Psi _{\mathrm{G}}\rangle =\Lambda _{h}\left( \sum_{ij}g_{ij}c_{i\uparrow
}c_{j\downarrow }\right) ^{\frac{N_{h}}{2}}|\mathrm{RVB}\rangle .  \label{gs}
\end{equation}%
Namely, the doped holes can be introduced by annihilating $N_{h}$ electrons
from the \textquotedblleft vacuum\textquotedblright\ state $|\mathrm{RVB}%
\rangle $. Apparently such a new state automatically satisfies the no double
occupancy constraint without invoking the Gutzwiller projection $\hat{P}_{%
\mathrm{G}}$. Furthermore, the holes are created in pairs with an amplitude $%
g_{ij}$ to realize the Cooper pairing, which is energetically favorable so
long as the RVB pairing persists in $|\mathrm{RVB}\rangle $ at low doping
\cite{weng_11}, as illustrated in Fig. \ref{wf}.

Here both $g_{ij}$ and $W_{ij}$ can be determined variationally \cite%
{weng_11} based on the t-J model and the main results are as follows. While $%
g_{ij}$ is associated with the singlet d-wave Cooper pairing, $W_{ij}$ is
found to change from a long-range power-law behavior ($\propto 1/\left\vert
\mathbf{r}_{ij}\right\vert ^{3}$) to an exponential behavior: $\left\vert
W_{ij}\right\vert \propto e^{-\frac{|\mathbf{r}_{ij}|^{2}}{2\xi ^{2}}}$,
where $\mathbf{r}_{ij}$ is the spatial distance and $\xi $ is the
characteristic pair size determined by the doping concentration $\delta $: $%
\xi =a\sqrt{\frac{2}{\pi \delta }}$ ($a$ is the lattice constant). This
indicates that $|\mathrm{RVB}\rangle $ indeed evolves from a long-range AFM
state to a \emph{short-ranged} spin liquid self-consistently at finite
doping.

Generally speaking, one cannot smoothly connect the ground state $|\Psi _{%
\mathrm{G}}\rangle $ in Eq. (\ref{gs}) to the Gutzwiller-projected BCS state
$|\Psi _{\mathrm{RVB}}\rangle $ in Eq. (\ref{BCS}). At finite doping, the
former is distinct from the latter by an explicit separation of the neutral
spin RVB pairing and Cooper pairing. Thus, in contrast to the
\textquotedblleft one-component\textquotedblright\ RVB state $|\Psi _{%
\mathrm{RVB}}\rangle $ without explicitly distinguishing the Cooper and RVB
pairings, the construction in Eq. (\ref{gs}) may be regarded as a
\textquotedblleft two-component\textquotedblright\ RVB structure. Such a
ground state can accurately describe the AFM correlations, via a bosonic
state $|\mathrm{RVB}\rangle ,$ at half-filling, while it can account for the
right charge degree of freedom at finite doping without an additional
projection procedure.

The key in driving an AFM state into a spin liquid involves an important
doping effect through $\Lambda _{h}$ in Eq. (\ref{gs}). Generally the phase
shift factor $\Lambda _{h}$ emerges because the bosonic RVB pairing in $|%
\mathrm{RVB}\rangle $ and the Cooper pairing of the electrons are not
statistically compatible and each doped hole will create a nonlocal phase
shift in the spin background. Here $\Lambda _{h}$ will represent the most
important and singular property of the Mottness as to be elaborated below.

\section{Mottness and sign structure}

The purely bosonic LDA wavefunction, Eq. (\ref{RVB}), can be regarded as an
important consequence of the Mott physics. Namely the electrons are fully
bosonized in the restricted Hilbert space at half-filling. In the opposite
dilute electron limit, the Fermi statistics of the electrons should get
recovered. Then the statistical sign structure of wavefunctions for a doped
Mott insulator is expected to be generally changed when holes are doped into
the system \cite{zaanen_09,weng_07}.

Explicitly identifying this effect turns out to be crucial in determining
the phase shift factor $\Lambda _{h}$ in the wavefunction construction (\ref%
{gs}), which may be generally expressed by
\begin{equation}
\Lambda _{h}\equiv \sum_{\{l_{h}\}}\left( n_{l_{1}}^{h}n_{l_{2}}^{h}\cdot
\cdot \cdot n_{l_{N_{h}}}^{h}\right) \varphi _{h}(l_{1},l_{2},\cdot \cdot
\cdot ,l_{N_{h}})e^{-i\left( \hat{\Omega}_{l_{1}}+\hat{\Omega}_{l_{2}}+\cdot
\cdot \cdot +\hat{\Omega}_{l_{N_{h}}}\right) }  \label{hphih}
\end{equation}%
where the phase shifts $\left\{ \hat{\Omega}_{l_{h}}\right\} $ are
associated with the holes, with $n_{l}^{h}=1-\sum_{\sigma }c_{l\sigma
}^{\dagger }c_{l\sigma }\geq 0$ denoting the hole occupation number at site $%
l$, and $\varphi _{h}$ is a bosonic wavefunction symmetric with regard to
the hole coordinates $\{l_{h}\}=l_{1},l_{2},\cdot \cdot \cdot ,l_{N_{h}}$,
which is present to ensure gauge invariance of the phase shift fields.

\begin{figure}[htbp]
\begin{center}
\subfigure[]{\includegraphics[width=100pt]{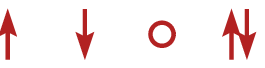}}\qquad\qquad\qquad %
\subfigure[]{\includegraphics[width=100pt]{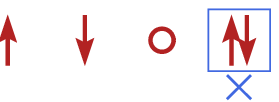}}
\end{center}
\caption{(a) Four possible electron states at a lattice site; (b) The double
occupancy state is forbidden under the Mottness constraint (\protect\ref%
{mottness}).}
\label{Hilbert}
\end{figure}

\begin{figure}[tbph]
\begin{center}
\includegraphics[width=300pt]{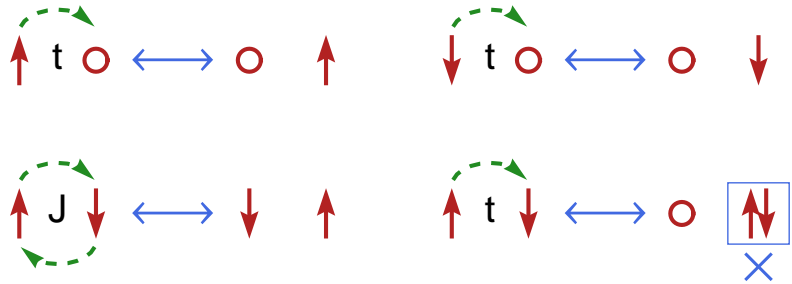}
\end{center}
\caption{The first three basic processes are allowed in the t-J model,
involving the nearest-neighboring spin-hole states.}
\label{t-J}
\end{figure}

Let us first introduce an exact theorem for the t-J model, known as the
phase string effect \cite{sheng_96,weng_97,WWZ_08}, which holds for a
bipartite lattice in any dimension, doping and temperature. The t-J model is
one of the simplest models describing doped Mott insulators in the limit
that the on-site Coulomb repulsion is much larger than the hopping integral $%
t$. This model is composed of two terms, $H_{t-J}=H_{t}+H_{J}$, with the
hopping term%
\begin{equation*}
H_{t}=-t\sum_{<ij>\sigma }c_{i\sigma }^{\dagger }c_{j\sigma }+H.c.
\end{equation*}%
and the superexchange term%
\begin{equation*}
H_{J}=J\sum_{<ij>}\left( \mathbf{S}_{i}\mathbf{\cdot S}_{j}-\frac{n_{i}n_{j}%
}{4}\right)
\end{equation*}%
where $\mathbf{S}_{i}$\textbf{\ }and $n_{i}\equiv \sum\nolimits_{\sigma
}n_{i\sigma }$ are on-site spin and number operators, respectively, and the
Hilbert space is constrained by Eq. (\ref{mottness}) as illustrated by Fig. %
\ref{Hilbert} and the basic processes are shown in Fig. \ref{t-J}.

\begin{figure}[bp]
\begin{center}
\subfigure[]{\includegraphics[width=170pt]{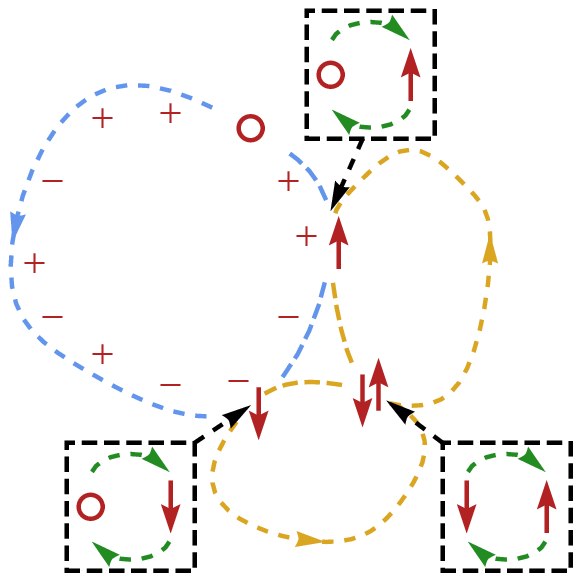}}\qquad\qquad\qquad %
\subfigure[]{\includegraphics[width=100pt]{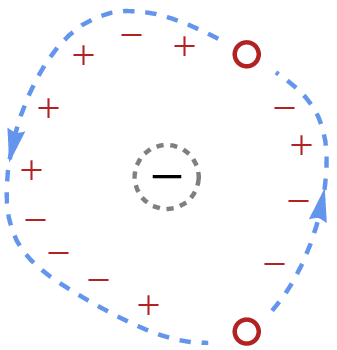}}
\end{center}
\caption{(a) A phase string associated with a typical closed path of a hole
moving on the spin background. Note that the $\pm $ signs are only created
by the exchanges between the hole and spins, via $\protect\tau _{c}$ in Eq. (%
\protect\ref{pstring1}), while pure spin loops do not contribute to any
signs. (b) The exchange of two holes in a close loop will lead to an
additional negative sign to $\protect\tau _{c}$. }
\label{loop}
\end{figure}

It can be rigorously demonstrated \cite{WWZ_08} that the partition function $%
Z_{t-J}=$Tr$\{e^{-\beta H_{t-J}}\}$ ($\beta =1/k_{\mathrm{B}}T$) in the
reduced Hilbert space (Figs. \ref{Hilbert} and \ref{t-J}) can be explicitly
expressed in terms of a loop summation%
\begin{equation}
Z_{t-J}=\sum_{c}\tau _{c}\mathcal{Z}(c)  \label{partition}
\end{equation}%
where $c$ denotes a set of multi-loops of holes/spins at arbitrary
temperature (for example, see Fig. \ref{loop}). Here $\mathcal{Z}(c)\geq 0$
is the positive weight for each closed path $c$ which depends on $t$, $J\,\ $%
and $\beta $ \cite{WWZ_08}.

Then $\tau _{c}$ collects all the nontrivial signs associated with the path $%
c$ as follows%
\begin{equation}
\tau _{c}=(-1)^{N_{h}^{\downarrow }(c)}\times (-1)^{N_{h}^{h}(c)}
\label{pstring1}
\end{equation}%
where $(-1)^{N_{h}^{\downarrow }(c)}$ is the so-called phase string with $%
N_{h}^{\downarrow }(c)$ counting the total number of the $\downarrow $-spins
that the holes have \textquotedblleft exchanged\textquotedblright\ with
along the loop $c$ as illustrated in Fig. \ref{loop}, which was first
discovered \cite{sheng_96} in the one-hole case of the t-J model. Physically
it describes the quantum string-like spin mismatches or the disordered
Marshall signs, created by the hopping of a hole in the spin background and
being dynamically irreparable. At finite doping, it remains irreparable as $%
\mathcal{Z}(c)\geq 0$, with an additional sign contribution by $%
(-1)^{N_{h}^{h}(c)}$ where $N_{h}^{h}(c)$ counts the total number of
hole-hole exchanges on the path $c$ (cf. Fig. \ref{loop}(b)).

Note that at half-filling, $\tau _{c}=1$ and there is no nontrivial sign in
the model, which is totally bosonized as pointed out above. In essence,
although the original electrons are fermions, due to the no double occupancy
constraint (\ref{mottness}), the Fermi-Dirac statistics will no longer play
a role in the restricted Hilbert space as far as the real physical process
is concerned. At finite doping, the nontrivial sign does reemerge in Eq. (%
\ref{partition}) as represented by Eq. (\ref{pstring1}). Although such sign
structure is solely associated with the doped holes and looks much
\textquotedblleft reduced\textquotedblright\ as compared to the full Fermi
signs associated with the original electrons, the phase string effect in $%
\tau _{c}$ will lead to the very singular effect: for any loop of a hole, a
mere change of one spin orientation can result in a total change of the sign
in $\tau _{c}$ and thus the phase interference from constructive to
destructive, or \emph{vice versa}. In particular, in contrast to the
conventional statistical problem involving the identical particles, here the
phase string effect is caused by the exchange between the holes and spins,
and so a mutual statistical effect will emerge \cite{weng_97}.
\begin{figure}[tph]
\begin{center}
\subfigure[]{\includegraphics[width=102pt]{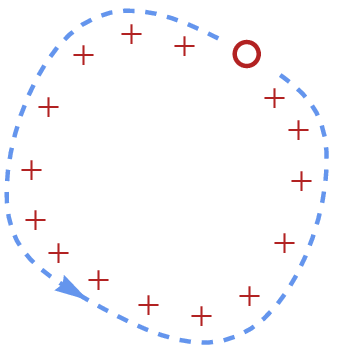}}\qquad\qquad\qquad %
\subfigure[]{\includegraphics[width=102pt]{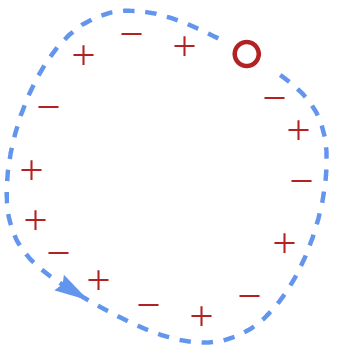}}\qquad\qquad\qquad %
\subfigure[]{\includegraphics[width=102pt]{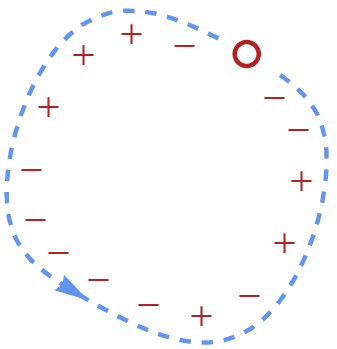}}
\end{center}
\caption{There will be no frustration from the phase string effect if a hole
is moving on the Nagaoka spin background where all the spins are polarized
along, say, the $\hat{z}$-direction as shown in (a). In (b), spins are in
the AFM ordered state, where even though the classical N\'{e}el order does
not lead to the negative total sign, the quantum spin fluctuations can
always result in a nontrivial phase string effect as one spin flip can cause
the total sign change. One expects the strongest phase string frustration on
the coherent motion of the hole if the spins are totally disordered as shown
in (c).}
\label{oneloop}
\end{figure}

\begin{figure}[bp]
\begin{center}
\includegraphics[width=180pt]{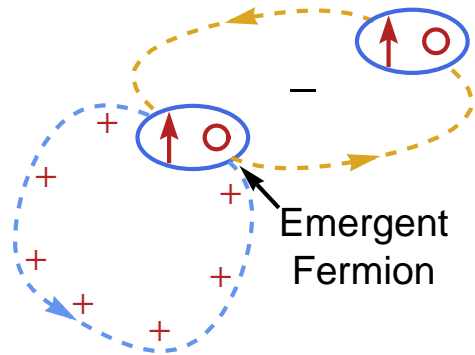}
\end{center}
\caption{Coherent fermions may emerge as tightly bound states of hole-spin
where a hole hopping always involves the same spin. Note that the bound
states are fermionic because of the additional sign upon the exchange of two
of them according to Fig. \protect\ref{loop}(b).}
\label{fermion}
\end{figure}

Figure \ref{oneloop} illustrates some typical phase strings picked up by a
closed loop motion of a hole in a spin background which is in (a) the
Nagaoka state in which spins are all polarized along the $\hat{z}$-axis; (b)
an AFM state with spins forming a N\'{e}el order; (c) a spin disordered
state. By noting that a closed path of the hole always involves an even
number of steps in a bipartite lattice, one can easily see that only in (a)
the phase interference due to phase string effect is always constructive,
whereas in (b) and (c) the phase string effect becomes nontrivial and
singular since even one spin flip can result in a total sign change of $%
(-1)^{N_{h}^{\downarrow }(c)}$. But in the partition function, the
probability for (a) is extremely rare at finite $J$ and thus the phase
string effect is generally crucial in mediating the \textquotedblleft
entanglement\textquotedblright\ between the spin and charge degrees of
freedom at finite doping. Figure \ref{fermion} further shows how fermionic
objects may arise from the bound pairs of holes and spins, which are free of
the phase string effect but retain the fermion signs between themselves via $%
(-1)^{N_{h}^{h}(c)}$ in Eq. (\ref{pstring1}). These fermions are associated
with the holes and represent only a small portion of some very specially
arranged loops in the summation of Eq. (\ref{partition}). In general, the
electrons are fractionalized as shown in Fig. \ref{loop} and the holes and
spins must obey a mutual statistics as dictated by the phase string effect.

Now one can introduce the precise definition of $\hat{\Omega}_{i}$ to
capture this mutual statistical effect. Define

\begin{equation}
e^{-i\hat{\Omega}_{i}}=e^{-\frac{i}{2}\left( \Phi _{i}^{s}-\Phi
_{i}^{0}\right) }\text{ ,}  \label{phif}
\end{equation}%
in which
\begin{equation}
\Phi _{i}^{s}\equiv \sum_{l\neq i}\theta _{i}(l)\left( \sum_{\sigma }\sigma
n_{l\sigma }^{b}\right) ~,  \label{phis}
\end{equation}%
and%
\begin{equation}
\Phi _{i}^{0}\equiv \sum_{l\neq i}\theta _{i}(l)\text{ ,}~  \label{phi0}
\end{equation}%
where $\theta _{i}(l)=\mathrm{Im}\ln $ $(z_{i}-z_{l})$ ($z_{i}$ is the
complex coordinate of site $i$), and $n_{l\sigma }^{b}$ denotes the spin
occupation number (with index $\sigma )$ at site $l,$ which always satisfies
the single occupancy constraint
\begin{equation}
\text{ }\sum_{\sigma }n_{l\sigma }^{b}=1\text{\ }  \label{constraint=1}
\end{equation}%
acting on the insulating spin state $|\mathrm{RVB}\rangle $.

\begin{figure}[tbph]
\begin{center}
\subfigure[]{\includegraphics[width=158pt]{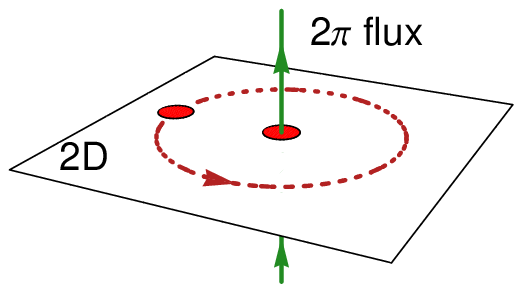}} %
\subfigure[]{\includegraphics[width=158pt]{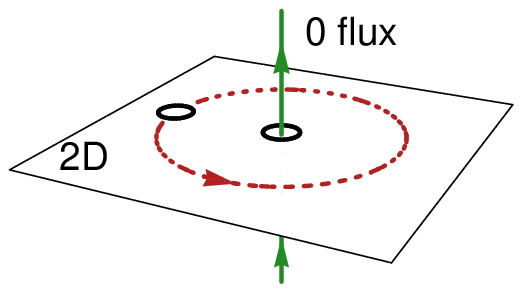}}\\[0pt]
\subfigure[]{\includegraphics[width=300pt]{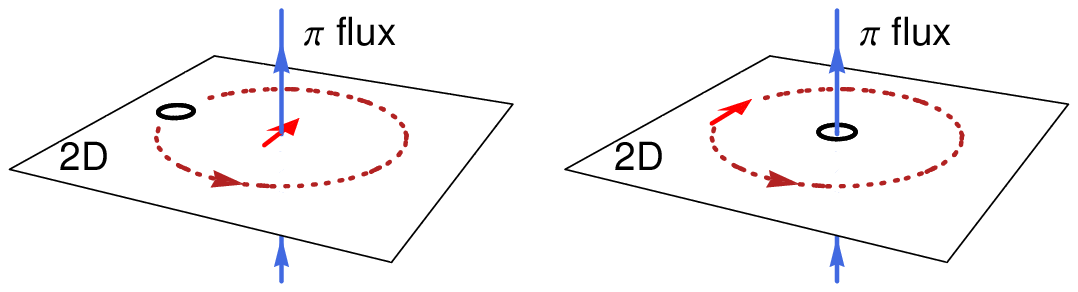}}
\end{center}
\caption{Statistical transmutations by using the flux attachment based on
the boson representation. (a) Fermionic statistics; (b) Bosonic statistics;
(c) Mutual semionic statistics.}
\label{statistics}
\end{figure}

In terms of Eq. (\ref{phif}), each spin in $|\mathrm{RVB}\rangle $ will
contribute a $\pm \pi $ vortex via $\Phi _{i}^{s}/2$ to the doped hole at
site $i$, with the spin itself sitting at the vortex core (see Fig. \ref%
{stat-transm}). \emph{Vice versa, }each doped hole will be perceived by the
spins in $|\mathrm{RVB}\rangle $ as introducing a $\pi $ vortex, also via $%
\Phi _{i}^{s}/2$, with the hole sitting at the core. It implies that a doped
hole and a neutral spin satisfy a \textquotedblleft mutual semion
statistics\textquotedblright\ as the phase shift $\hat{\Omega}_{i}$ amounts
to giving rise to $\pm \pi $ when one kind of species continuously circles
around the other one once as illustrated in Fig. \ref{statistics}(c). Note
that the single-valueness of Eq. (\ref{phif}) will be ensured by combining
with $\Phi _{i}^{0}/2$. Thus the total phase shift added up in $\Lambda _{h}$
represents a nontrivial entanglement between the doped holes and background
spins, which will decide a \textquotedblleft mutual semion
statistics\textquotedblright\ sign structure in $|\Psi _{\mathrm{G}}\rangle $
that is fundamentally different from that in a BCS state $|\mathrm{BCS}%
\rangle $ satisfying the Fermi-Dirac statistics (cf. Fig. \ref{statistics}).
\begin{figure}[tbph]
\begin{center}
\includegraphics[width=280pt]{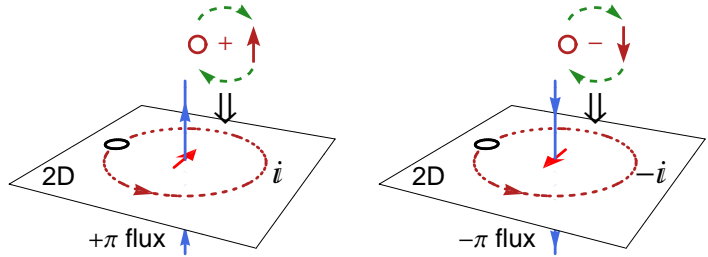}
\end{center}
\caption{The phase string effect of the t-J model can be precisely realized
via the phase shift factor $\Lambda _{h}$ in the wavefunction (\protect\ref%
{wf}).}
\label{stat-transm}
\end{figure}

One may further examine such a sign structure by a thinking experiment in
which a hole in $|\Psi _{\mathrm{G}}\rangle $ goes through a closed loop $c.$
At each step of nearest-neighbor moving of the hole, a singular phase $0$ or
$\pi $ is generated via a phase shift $\hat{\Omega}_{i}$ in $\Lambda _{h}$
depending on $\uparrow $ or $\downarrow $ spin that the hole
\textquotedblleft exchanges\textquotedblright\ with as shown in Fig. \ref%
{stat-transm}. Although $\Lambda _{h}$ also produces other phase shift
contributed by other spins not \textquotedblleft
exchanged\textquotedblright\ with the hole, but their net effect is zero
after summing up the total phase for a closed-path motion of the hole. In
the end, one finds
\begin{equation}
|\Psi _{\mathrm{G}}\rangle \rightarrow (-1)^{N_{h}^{\downarrow }(c)}|\Psi _{%
\mathrm{G}}\rangle   \label{pstring}
\end{equation}%
to reproduce the phase string sign factor. It is straightforward to verify
that that fermionic signs of the doped holes created by $\left(
\sum_{ij}g_{ij}c_{i\uparrow }c_{j\downarrow }\right) ^{\frac{N_{h}}{2}}$ in
Eq. (\ref{gs}) can precisely account for the additional sign factor $%
(-1)^{N_{h}^{h}}$ in Eq. (\ref{pstring1}). Consequently, the nontrivial sign
structure $\tau _{c}$ identified in the t-J model is naturally incorporated
into the ground state $|\Psi _{\mathrm{G}}\rangle $ in Eq. (\ref{gs}) since
the bosonic spin background $|\mathrm{RVB}\rangle $ does not contribute to
any additional statistical signs.

Therefore, comparing with the Gutzwiller-projected BCS state (\ref{BCS}),
one finds that the new ground state (\ref{gs}) has the following distinct
and unique features. First, it recovers the LDA ground state at
half-filling, which is known as the most accurate variational wavefunction
for the Heisenberg antiferromagnet. The Fermi statistics completely
disappears in this limit, and the wavefunction satisfies the well-known
Marshall signs. Second, at finite doping, the new ground state (\ref{gs})
satisfies the mutual statistics via the phase shift field $\hat{\Omega}_{i}$
defined in Eq. (\ref{phif}), which reproduces the precise sign structure
identified in the t-J model. The form of the wavefunction is therefore
generally fixed by these two unique requirements derived from the t-J model,
and the adjustable parameters $\varphi _{h}$, $g_{ij}$, and $W_{ij}$ are
presumably smooth functions. For a uniform solution with $\varphi _{h}=$
constant, a d-wave $g_{ij}$ with a short-ranged $W_{ij}$ can be determined
variationally based on the t-J model as mentioned in the previous section,
in which the phase shift field in $\Lambda _{h}$ plays a central role in
turning the long-range RVB pairing into the short-range one at finite doping
\cite{weng_11}. In the following we illustrate how the superconducting phase
coherence can be simultaneously realized once the AFM long-range order in $|%
\mathrm{RVB}\rangle $ is turned into a spin liquid.

\section{Superconducting phase coherence}

Generally speaking, the Cooper pairing amplitude is already preformed in Eq.
(\ref{gs}) (with the pairing symmetry determined by $g_{ij}$), but the true
superconducting off-diagonal-long-range-order (ODLRO) will be determined by
the phase coherence through $\hat{\Omega}_{i}$ in $\Lambda _{h}$, which
sensitively depends on the spin correlation in $|\mathrm{RVB}\rangle $. Now
we examine the condition under which the ground state has a true
superconducting ODLRO.

With the pre-existence of the Cooper pairing amplitude in Eq. (\ref{gs}),
the superconducting order will be determined by \cite{weng_11}
\begin{equation}
\left\langle c_{i\uparrow }c_{j\downarrow }\right\rangle \propto
\left\langle \mathrm{RVB}\right\vert e^{i\left( \hat{\Omega}_{i}+\hat{\Omega}%
_{j}\right) }|\mathrm{RVB}\rangle .  \label{phcoh-0}
\end{equation}%
We further note that, due to the presence of $\Lambda _{h}$, injecting a
hole into the ground state $|\Psi _{\mathrm{G}}(N_{h})\rangle $ in Eq. (\ref%
{gs}) will also induce a phase shift by
\begin{equation}
c_{i\sigma }|\Psi _{\mathrm{G}}(N_{h})\rangle \sim e^{i\hat{\Omega}%
_{i}}|\Psi _{\mathrm{G}}(N_{h}+1)\rangle .  \label{pshift}
\end{equation}%
Namely the wavefunction overlap between the bare hole state and the true
ground state of $N_{h}+1$ holes will depend on $e^{i\hat{\Omega}_{i}}$.
Hence the phase coherence of superconductivity and the coherence of a Landau
(or more precisely, Bogoliubov) quasiparticle will be simultaneously
realized. And a \textquotedblleft normal state\textquotedblright\ obtained
by disordering the phase shift factor $e^{i\hat{\Omega}_{i}}$ will be
intrinsically a non Fermi liquid with vanishing quasiparticle weight.

If a long-range RVB pairing (i.e., $W_{ij}$ is in a power-law decay at large
$\left\vert \mathbf{r}_{ij}\right\vert $) is present, like in the AFM
ordered phase, the phase coherence in Eq. (\ref{phcoh-0}) will be generally
destroyed because the $\pm \pi $ vortices in $\hat{\Omega}_{i}$ carried by
the two spinon partners of an RVB pair, according to Eq. (\ref{phif}), do
not compensate each other, which results in phase disordering. Only can a
short-ranged RVB pairing lead to a vortex-antivortex binding in Eq. (\ref%
{phcoh-0}) and thus the superconducting phase coherence. In other words, in
the superconducting phase $|\mathrm{RVB}\rangle $ has to become a spin
liquid with short-range AFM\ correlations. As previously mentioned, the RVB
amplitude $W_{ij}$ does become short-ranged with a finite, doping-dependent $%
\xi $ based on a self-consistent calculation in terms of Eq. (\ref{gs}).

\begin{figure*}[bph]
\centering
\includegraphics[width=5cm]{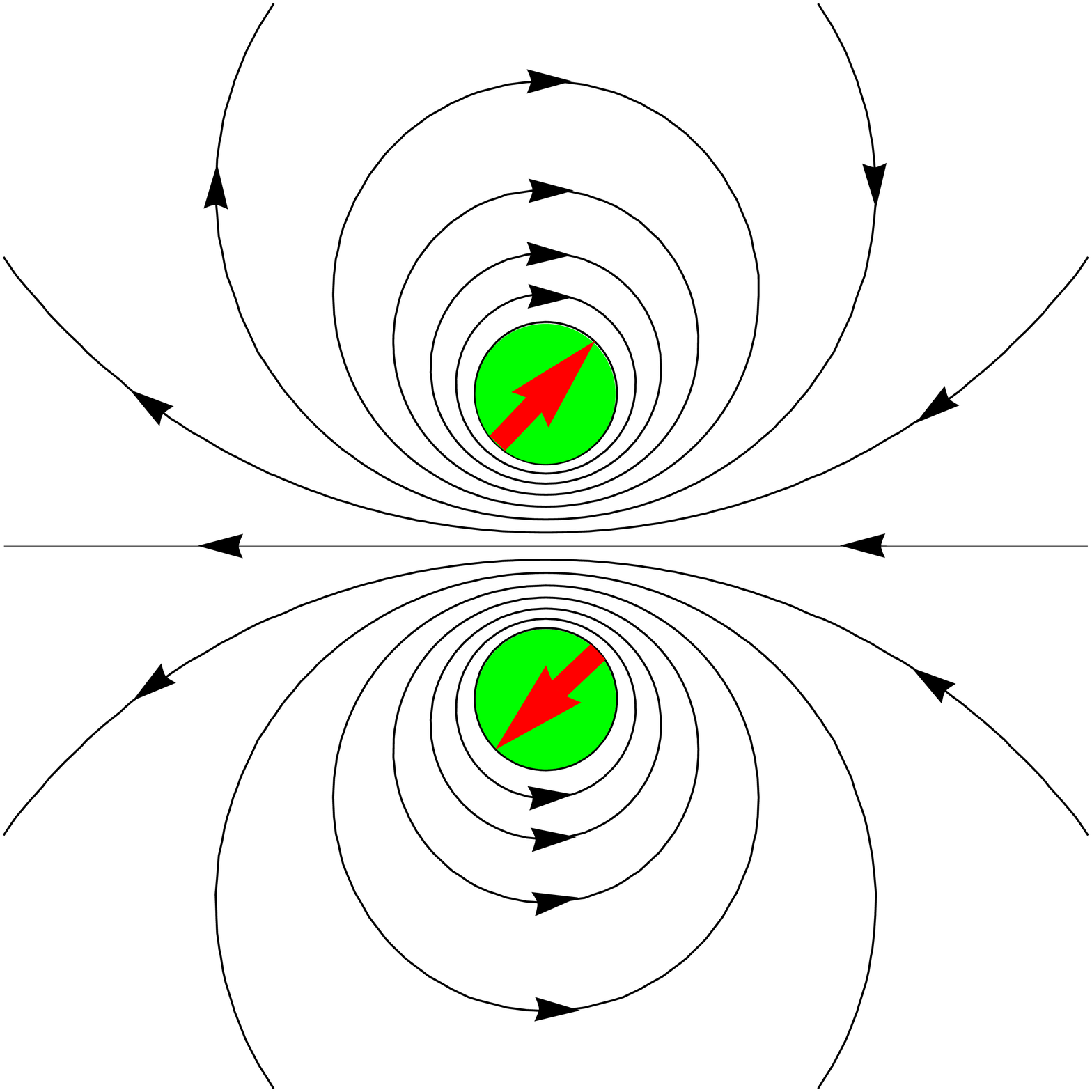}\qquad %
\includegraphics[width=5cm]{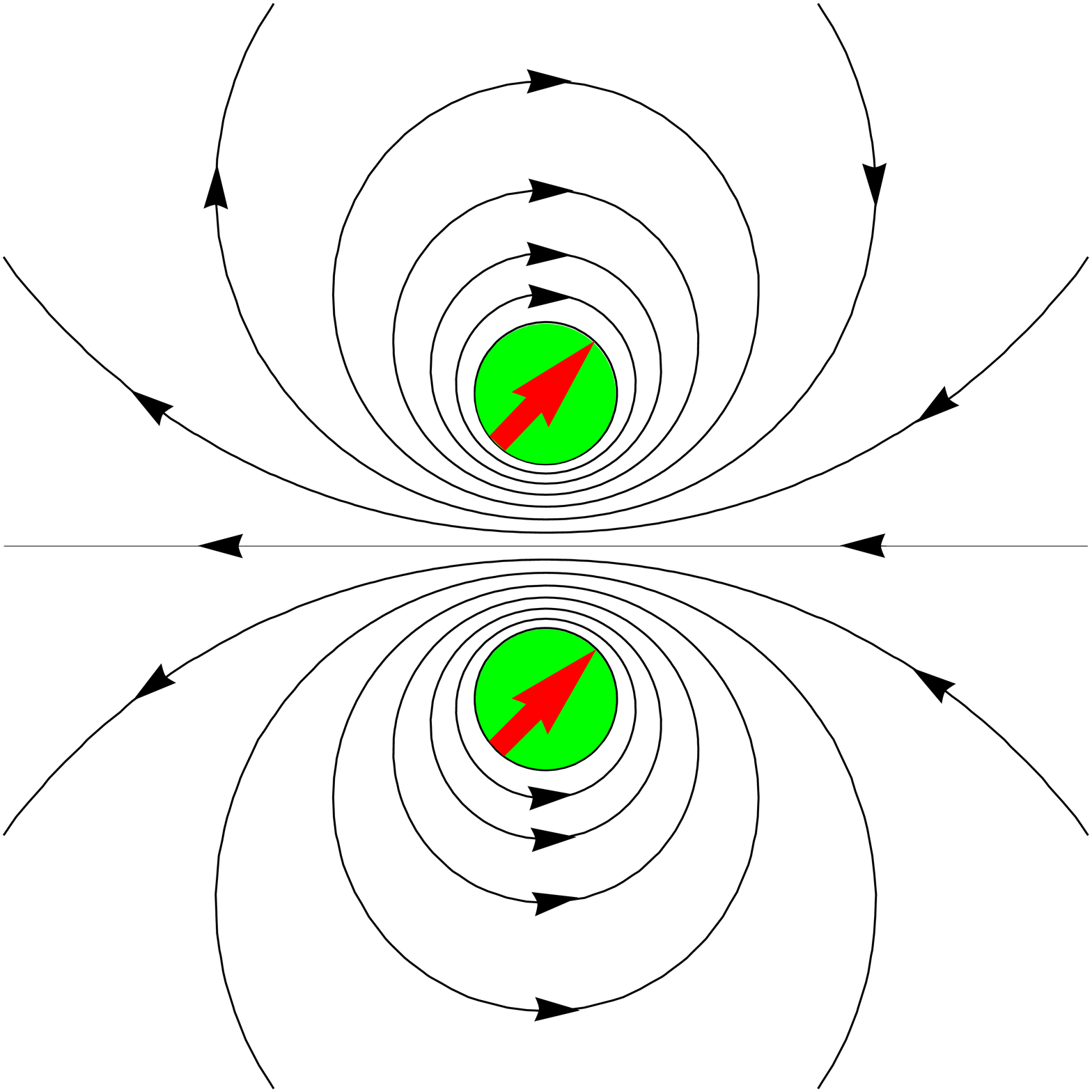}
\caption{Spin-rotons with $S=0$ and $S=1$, respectively, which are
identified \protect\cite{MW_10} as the most essential elementary excitations
on top of the ground state (\protect\ref{gs}) and decide the $T_{c}$-formula
(\protect\ref{tc}). }
\label{spin-rotons}
\end{figure*}

Furthermore, the statistical sign structure $\Lambda _{h},$ which should be
generally present in the excited states of the t-J model as well, will play
a critical role to control the superconducting phase transition. Imagine a
pair of spinon excitations created by breaking up an RVB pair in $|\mathrm{%
RVB}\rangle $, which through the phase shift in $\Lambda _{h}$ will lead to
two $\pm \pi $ vortices according to Eq. (\ref{phif}). Normally the free
vortices would disorder the superconducting phase coherence, and in order to
maintain the latter below $T_{c}$, a pair of excited spinons will be forced
to form a vortex and antivortex bound pair with the spinons sitting at the
vortex cores, which are known as the spin-rotons \cite{MW_10} as illustrated
in Fig. \ref{spin-rotons}. In contrast to the singlet RVB pair in the ground
state, a spin-roton is a pair of loosely bound (confined) excited spinons
via a logarithmic potential with the spin quantum numbers, $S=0$ and $1$
\cite{MW_10}.

As detailed in Ref. \cite{MW_10}, spin-roton excitations will eventually
destroy the phase coherence at $T_{c}$, and a simple $T_{c}$ formula has
been obtained as follows \
\begin{equation}
T_{c}\simeq \frac{E_{g}}{6k_{\mathrm{B}}}  \label{tc}
\end{equation}%
with $E_{g}\sim \delta J$ denoting the core energy of the spin-rotons,
degenerate for $S=0$ and $1$. This relation is in excellent agreement with
the empirical formula established in the experiment\cite{Uemura}, with $S=0$
and $S=1$ spin-rotons corresponding to the \textquotedblleft
resonancelike\textquotedblright\ modes observed in the Raman $A_{\mathrm{1g}}
$ channel and neutron scattering measurements, respectively. It also
provides a natural explanation of why the two modes are energetically
degenerate in the experiment\cite{Uemura}.

\begin{figure}[bp]
\centering \includegraphics[width=5cm]{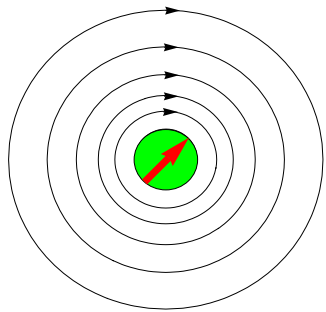}
\caption{A spinon-vortex is a composite object with an $S=1/2$ free spin
locking with a supercurrent vortex. It automatically appears in Eq. (\protect
\ref{gs}) if a single spinon is present in $|\mathrm{RVB}\rangle $. A
spin-roton excitation in Fig. \protect\ref{spin-rotons} can be regarded as a
vortex-antivortex bound pair of the spinon-vortices.}
\label{spinon-vortex}
\end{figure}

Here the spin-rotons are the most essential elementary excitations of
non-BCS-type in a superconductor of doped Mott insulators characterized by
Eq. (\ref{gs}), which directly controls the superconducting phase coherence
in Eq. (\ref{phcoh-0}) via a characteristic energy $E_{g}$. At $T>T_{c}$,
spin-roton excitations (with $S=0,1$) will disassociate into free
spinon-vortices ($S=1/2$) as shown in Fig. \ref{spinon-vortex}. In the AFM
long-range ordered state near half-filling, one has $E_{g}\rightarrow 0$
such that $T_{c}$ vanishes, and the $S=1$ spin-roton excitation will
naturally reduce to the gapless spin wave with the RVB pairing becoming a
long-ranged one. These non-superconducting states will be further discussed
below.

\section{Nature of pseudogap states}

Lying between the AFM phase and superconducting phase, there generally
exists a crossover region in the phase diagram of the cuprates, which
exhibits the so-called pseudogap phenomenon. Such a region is so unique to
the cuprates and has been heavily investigated experimentally \cite%
{timusk_99,shen_03} and theoretically \cite{LNW_06,weng_07} in hoping to
crack the high-$T_{c}$ mystery.

The present superconducting ground state (\ref{gs}) is constructed using two
criteria: It satisfies the general statistical sign structure of the t-J
model and correctly reduces to the AFM state at half-filling. Therefore it
can provide, in principle, an unequivocal prediction for the crossover
region connecting the AFM and superconducting regimes.

By definition, a pseudogap state should be a non-superconducting version of
Eq. (\ref{gs}) at finite doping. Indeed, the structure of Eq. (\ref{gs})
does suggest a series of non-superconducting \textquotedblleft ground
states\textquotedblright \ as outlined below.

Pseudogap Region I. As already pointed out in the previous section, at $%
T\gtrsim T_{c}$, spin-roton excitations will disassociate into free
spinon-vortices (cf. Fig. \ref{spinon-vortex}) to disorder the
superconducting phase coherence. The resulting vortex liquid state \cite%
{WM_02,WQ_06,weng_07} provides a microscopic description of the Nernst
regime discovered \cite{nernst} in the cuprates. Now to bring such a vortex
liquid state to $T=0$, one has to make $E_{g}\rightarrow 0$ in Eq. (\ref{tc}%
) or equivalently the RVB pairing in $|\mathrm{RVB}\rangle $ go to a
long-ranged one to result in the right-hand-side of Eq. (\ref{phcoh-0})
vanishing. This is a quantum vortex ground state, in which the $d$-wave
Cooper pair amplitude can be still maintained via $g_{ij}$, but the Cooper
pairs apparently become incoherent because of phase disordering. On the
other hand, although $W_{ij}$ is in a power-law decay in $|\mathrm{RVB}%
\rangle $, the long-range AFM order is also disordered by $\Lambda _{h}$ via
the same phase shift fields due to the presence of doped holes in
(incoherent) Cooper pairing. Such a peculiar \textquotedblleft mutual
duality\textquotedblright\ phase is called Bose insulator phase in Ref. \cite%
{YTQW_11}, where the statistical sign structure in $\Lambda _{h}$ has been
formulated in terms of a mutual Chern-Simons gauge theory description.

In literature, a vortex liquid state has been usually considered \cite%
{kivelson,LNW_06,tesanovic_08} as the result of phase fluctuations in the
charge-2e d-wave pairing order parameter at low superfluid density.
Pseudogap Region I apparently shares this similarity according to Eq. (\ref%
{phcoh-0}). However, I stress here that the electron fractionalization and
mutual statistics further specify that each vortex has always a free $S=1/2$
spinon sitting at the vortex core (cf. Fig. \ref{spinon-vortex}), and the
charge-2e pairing order parameter is just a composite quantity involving
both RVB and Cooper channels \cite{weng_11}. Namely the spin degrees of
freedom control the charge behavior, including $T_{c}$ (Eq. (\ref{tc})) as
well as the Nernst properties \cite{WM_02,WQ_06,weng_07}. It is a unique
characteristic of the mutual statistics that is clearly distinct from a
conventional vortex liquid state.

Pseudogap Region II. One may obtain another non-superconducting ground state
in Eq. (\ref{gs}) in the case that $|\mathrm{RVB}\rangle $ has become a
short-ranged spin liquid with a finite $E_{g}$. In this case, the weaker
Cooper pair amplitude may be made to disappear in $g_{ij}$, say, via strong
magnetic fields, while the Zeeman energy still remains small as compared to $%
E_{g}$ to let $|\mathrm{RVB}\rangle $ intact. Then in such a pseudogap
state, a small Fermi pocket will emerge in Eq. (\ref{gs}) with the doped
holes behave like free fermions moving in the vacuum of the spin liquid $|%
\mathrm{RVB}\rangle $. Because of the gap $E_{g}$ in $|\mathrm{RVB}\rangle $%
, the phase coherence is still maintained in $\Lambda _{h}$ at $T\ll E_{g}$
such that\ the small coherent Fermi pocket is expected to contribute to the
quantum oscillation effect without involving a translational symmetry
breaking. Here the pseudogap behavior will be mainly exhibited in $|\mathrm{%
RVB}\rangle $, which may be probed by the NMR experiments as discussed
recently in Ref. \cite{Mei_11}.

It is noted that the state of Pseudogap Region II is similar to the dopon
theory \cite{wen_05} in which small Fermi pockets can also be induced \cite%
{Mei_11} by strong magnetic fields. The main difference lies in the
short-range RVB nature of $|\mathrm{RVB}\rangle $, which is different from a
nodal fermion spin liquid state in the dopon theory \cite{wen_05,Mei_11}. On
the other hand, a short-range bosonic spin liquid background with small
Fermi pockets has been phenomenologically discussed in Ref. \cite{Qi Yang_10}
recently.
\begin{figure}[bph]
\begin{center}
\subfigure[]{\includegraphics[width=100pt]{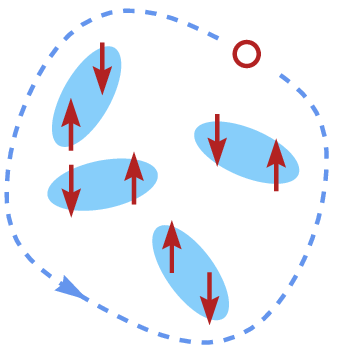}}\qquad\qquad\qquad %
\subfigure[]{\includegraphics[width=100pt]{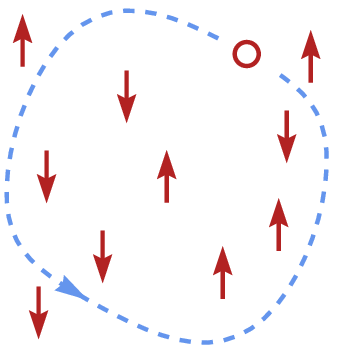}}
\end{center}
\caption{(a): The motion of the doped hole is coherent in a short-range RVB
background; (b): The strongest scattering between the hole and spin degrees
of freedom occurs when the background spins are uncorrelated in the
high-temperature \textquotedblleft strange metal phase\textquotedblright ,
where each spin carries a fictitious $\pm \protect\pi $ fluxoids as
perceived by the hole (cf. also Fig. \protect\ref{oneloop}(c)). }
\label{max}
\end{figure}

Note that these pseudogap states generalized from Eq. (\ref{gs}) are not
necessarily associated with any explicit symmetry-breaking orders. They
emerge as a necessary crossover from the AFM state at half-filling towards
the superconducting state at a finite doping. What is in common in these
states is the presence of AFM correlations in $|\mathrm{RVB}\rangle $ in the
form of RVB pairing. What differentiates Pseudogap Region I and II is
whether there is a further preformed Cooper pairing or not in $g_{ij}$.
Finally the pseudogap properties are gone if the RVB pairing disappears in $|%
\mathrm{RVB}\rangle $ at high temperatures, leading to a Curie-Weiss
classical spin background \cite{Gu_07}. The doped holes will get maximally
scattered via the phase string effect in $\Lambda _{h}$ as illustrated by
Fig. \ref{max}(b), in contrast to the more coherent motion in the RVB regime
(Fig. \ref{max}(a)). As shown in Ref. \cite{Gu_07}, the resistivity in this
regime will exhibit a linear-$T$ \textquotedblleft strange
metal\textquotedblright\ behavior.

\section{Concluding remarks}

In the prototype RVB state (\ref{BCS}), the Mottness simply implies the
Gutzwiller projection onto a Hilbert space that excludes double occupancy of
the electrons. Note that double occupancy is still allowed in $|\mathrm{BCS}%
\rangle $, which is basically a Fermi liquid with the Cooper channel
instability. A fundamental question is if such a Gutzwiller projection $\hat{%
P}_{\mathrm{G}}$ acting on a suitably chosen \textquotedblleft
ordinary\textquotedblright\ state $|\mathrm{BCS}\rangle $ (which may be
considered as the ground state of an effective \emph{local} Hamiltonian of
fermions) is sufficient to capture the essential physics of the Mottness?

Through this article I have pedagogically reviewed the efforts \cite%
{weng_07,weng_11,zaanen_09} to reexamine this important question. The basic
conclusion is that, as a matter of principle, $|\mathrm{BCS}\rangle $ cannot
be \textquotedblleft ordinary\textquotedblright\ as described above and the
singular \textquotedblleft feedback\textquotedblright\ effect of the
constrained Hilbert space should turn the effective Hamiltonian of $|\mathrm{%
BCS}\rangle $, if exists, into a new paradigm of non-Fermi-Dirac statistics.
In other words, we have reached the conclusion that the essential physics of
the Mottness is not simply about imposing the constraint (\ref{mottness}) on
the Hilbert space. Rather, an electron fractionalization with mutual
fractional statistics must take place in $|\mathrm{BCS}\rangle $ as a
general consequence of the Mottness \cite{zaanen_09,weng_97}.

The resulting new ground state \cite{weng_11} as given in Eq. (\ref{gs}) is
distinct from the original RVB state \cite{pwa_87} in Eq. (\ref{BCS}) by
that the neutral RVB and Cooper channels are clearly differentiated. It
implies an electron fractionalization in which three essential types of
correlations are explicitly separated and intrinsically embedded: i.e., the
AFM correlations in $|\mathrm{RVB}\rangle $; the Cooper pairing of the doped
holes; and the mutual influence/competition between these two channels via a
mutual statistical phase in $\Lambda _{h}$.

As emphasized in this article, an accurate description of the AFM state is a
crucial starting point. So does the singular doping effect, which brings in
the fundamental change in the statistical sign structure known as the phase
string effect \cite{sheng_96,weng_97,WWZ_08}. Based on these two basic
properties of the Mott physics, the pseudogap phenomenon and
superconductivity arise naturally as a self-organization of the three
essential types of correlations mentioned above. Without holes, the
long-range AFM order will always win in $|\mathrm{RVB}\rangle $, which
provides a highly accurate description \cite{lda_88} of the ground state of
the Heisenberg Hamiltonian. A sufficient concentration of doped holes will
eventually turn the antiferromagnetism in $|\mathrm{RVB}\rangle $ into a
true\ spin liquid state with short-range AFM correlations. By doing so the
Cooper paired holes can gain phase coherence to realize a high-$T_{c}$
superconductivity. The pseudogap region is merely a crossover from the
long-range AFM order to superconducting order since the latter two are
incompatible in nature. This incompatibility decides the complex crossover
phenomenon of the pseudogap physics \cite{weng_07}. There is nothing
mysterious about them once the essence of the Mottness is properly
understood.

Deep inside the superconducting state, the spins are short-range RVB paired
in $|\mathrm{RVB}\rangle $ such that the statistical signs get cancelled in $%
\Lambda _{h}$, and the state (\ref{gs}) resembles Eq. (\ref{BCS}) proposed
by Anderson \cite{pwa_87} in that the Cooper pairing is driven by the spin
RVB pairing upon doping. The d-wave pairing symmetry and the RVB state as a
prototype pseudogap phase were actually predicted \cite{zhang_88,bza_87}
before experimental discoveries. In fact, in a modified version of Eq. (\ref%
{BCS}), the \textquotedblleft fugacity\textquotedblright\ has been also
introduced to stress the tendency towards a separation of the neutral RVB
and Cooper pairings \cite{pwa_11}. Nevertheless, the statistical signs in $%
\Lambda _{h}$ of Eq. (\ref{gs}) will play a critical role in dictating how
the spin excitation affects the superconductivity and \emph{vice versa}. In
order to properly understand the superconducting phase transition, the
nature of pseudogap physics, and even the strange metal behavior at high
temperatures, the mutual statistics encoded in\ $\Lambda _{h}$ cannot be
omitted in a theory based on the doped Mott insulator approach.

\begin{acknowledgments}
I acknowledge useful discussions with P. W. Anderson, W.-Q. Chen, Z.-C. Gu,
S.-P Kou, T. Li. J.-W. Mei, V. N. Muthukumar, H. T. Nieh, N. P. Ong, X.-L.
Qi, Y. Qi, D.-N. Sheng, C.-S. Tian, Y.-Y. Wang, X.-G. Wen, K. Wu, P. Ye, J.
Zaanen, F. Zhou, and Y. Zhou. I am grateful to Y.-Z. You's help with the
figures. This work was supported by NSFC No. 10834003, National Program for
Basic Research of MOST grant nos. 2009CB929402 and 2010CB923003.
\end{acknowledgments}

\end{document}